\documentstyle[epsfig,12pt]{article}

\newcommand{\bce}{\begin{center}} 
\newcommand{\ece}{\end{center}}
\newcommand{\beq}{\begin{equation}}
\newcommand{\eeq}{\end{equation}}
\newcommand{\bea}{\vspace{0.25cm}\begin{eqnarray}}
\newcommand{\eea}{\end{eqnarray}}

\newcommand{\brho}{\mbox{\boldmath $\rho$}}

\newcommand{\bp}{{\bf p}}

\newcommand{\br}{{\bf r}}
\newcommand{\bR}{{\bf R}}

\newcommand{\ba}{\begin{array}}
\newcommand{\ea}{\end{array}}

\newcommand{\doublespace}{
    \renewcommand{\baselinestretch}{1.6}\large\normalsize}

\newcommand{\bfrho}{\mbox{\boldmath $\rho$}}

\newcommand{\bkappa}{\mbox{\boldmath ${\kappa}$}}

\def\lsim{\mathrel{\rlap{\lower4pt\hbox{\hskip1pt$\sim$}}
    \raise1pt\hbox{$<$}}}         
\def\gsim{\mathrel{\rlap{\lower4pt\hbox{\hskip1pt$\sim$}}
    \raise1pt\hbox{$>$}}}         

\def\Pom{{\bf I\!P}}

\def\beq{\begin{equation}}
\def\endeq{\end{equation}}
\def\arr{\begin{eqnarray}}
\def\endarr{\end{eqnarray}}
\makeindex

  \advance\oddsidemargin  by -1in
  \advance\evensidemargin by -1in
  \advance\topmargin      by -0.5in
\begin{document}

\vspace{2.0cm}

\begin{flushright}
ITEP-PH-4/2004\\
FZJ-IKP-TH-2004-18
\end{flushright}

\vspace{1.0cm}

\begin{center}
{\Large \bf 
Evolution of high-mass diffraction from the 
light quark valence component of the pomeron}

\vspace{1.0cm}

{\large\bf N.N.~Nikolaev$^{a,b}$, W. Sch\"afer$^a$,
B.G.~Zakharov$^{b}$ and V.R.~Zoller$^{c}$}

\vspace{1.0cm}
{\sl
$^{a}$IKP(Theorie), FZ J{\"u}lich, J{\"u}lich, Germany
\medskip\\
$^{b}$L.D. Landau Institute for Theoretical Physics, Moscow 117940, Russia
\medskip\\
$^{c}$ ITEP, Moscow 117218, Russia.
}\vspace{1.0cm}\\
{ \bf Abstract }\\
\end{center}

We analyze the contribution from excitation of
the $(q\bar q)(f\bar f),(q\bar q)g_1...g_n(f\bar f)$ 
Fock states
of the photon  to high mass diffraction in DIS.
We show that the large $Q^2$ behavior of this contribution can be described 
by the DLLA evolution from the non-perturbative $f\bar f$ valence state of 
the pomeron. Although  of higher order in pQCD, the new contribution to high-mass
diffraction is 
comparable to that from the excitation of the $q\bar q g$ Fock state of the 
photon.

\doublespace
\pagebreak



In this communication we report a direct evaluation of high-mass diffractive 
deep inelastic scattering (DIS) 
from excitation of the $(q\bar{q})(f\bar{f})$ Fock states of the photon,
fig. 1, 
where $q$ and $f$ are the light quarks. 
The interest
in this problem can be formulated as follows. 
\begin{figure}[h]
\psfig{figure=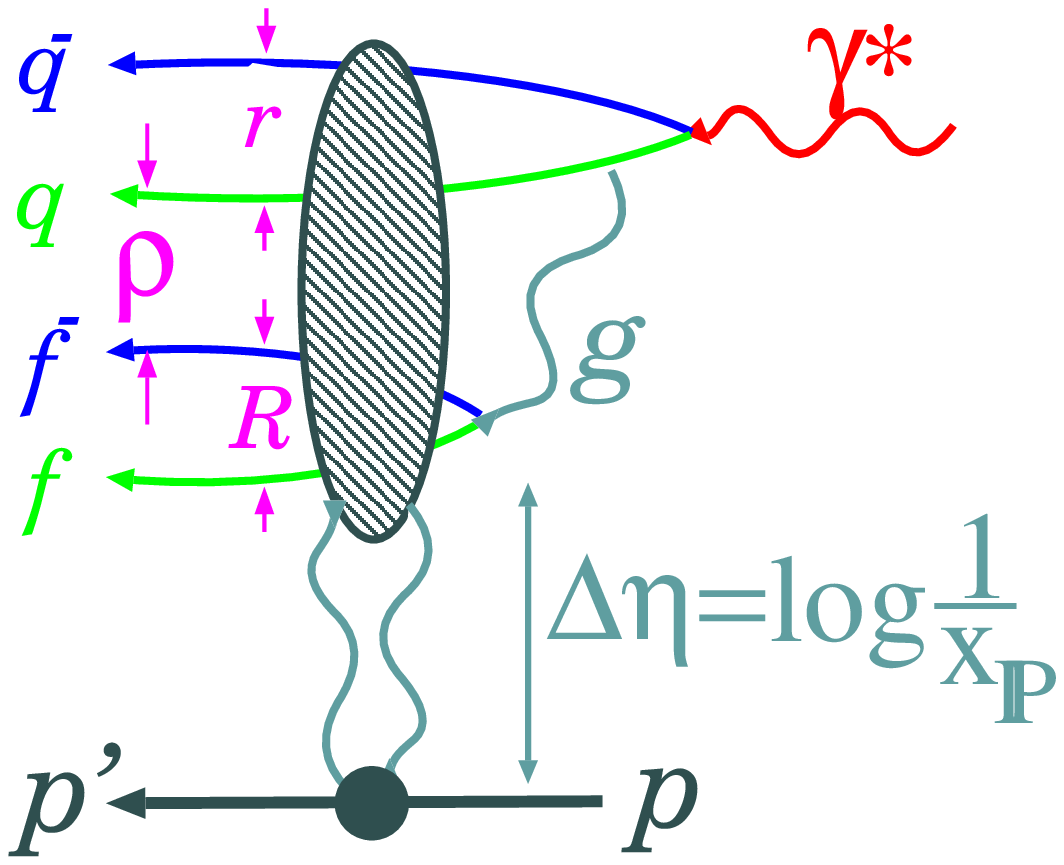,height=6cm}
\caption{\it The color dipole structure of diffractive 
excitation of perturbative $(q\bar{q})(f\bar{f})$
state of the photon with the rapidity gap $\Delta \eta$ from the proton. 
The contribution of other
diagrams with radiation of the gluon and/or splitting $g\to f\bar f$ after
the interaction with the target vanishes to DLLA.}
\end{figure}  

On the one hand, within perturbative QCD (pQCD) diffractive DIS,
$\gamma^*p \to X+p'$, can be
described as quasi-elastic scattering and excitation of the multiparton 
Fock states $X$ of the incident photon of virtuality $Q^2$ \cite{NZ92,NZ3P}.
As such it is a manifestly nonlinear - quadratic - functional of the
dipole cross section for the multiparton states,
$X= q\bar{q},q\bar{q}g,...$. For the forward case, $t=0$, where 
$t$ is the $(p,p')$ momentum transfer squared, 
\bea
\left.{\frac{d\sigma^D}{ dt}}\right|_{t=0}={\frac{1}{ 16\pi}}
\left[\langle q\bar{q}|\sigma^2|q\bar{q}\rangle
+\langle q\bar{q}g|\sigma_3^2-\sigma^2|q\bar{q}g\rangle+...\right] \, ,
\label{eq:DSDT}
\eea
where $\sigma$ and $\sigma_3$ stand for the dipole cross section for 
the Fock states $|q\bar{q}\rangle,
|q\bar{q}g\rangle$, respectively, interacting with the
 proton target, evaluated at
the starting point of the small-$x$ evolution, $x_{\Pom}=x_0$.
On the other hand, motivated by the triple-reggeon approach to diffraction
excitation \cite{KTM}, one would like to reinterpret high-mass diffractive DIS
as an inclusive DIS off the pomeron, 
\bea
(Q^{2}+M^{2})\left.{\frac{d\sigma^{D}} { dt dM^{2}}}\right|_{t=0}& = &
\langle q\bar{q}|\sigma^{\Pom}(x_{\Pom},\beta,\br)|q\bar{q}\rangle\,.
\label{eq:DSDT2}
\eea
If possible at all, such a color dipole representation will only be 
meaningful if the effect of higher Fock states of the photon can 
consistently be reabsorbed into the small-$\beta$ evolution 
of $\sigma^{\Pom}(x_{\Pom},\beta,\br)$ \cite{NZ3P,NZZ94}. Here 
$\beta=Q^2/(M^2+Q^2)$, where $M$ is the mass of the diffractive system,
is the Bjorken variable for DIS off the pomeron, $x_{\Pom}=x/\beta$
is the rapidity gap variable - the fraction of the proton's lightcone
momentum carried by the exchanged pomeron, $x=Q^2/2m\nu$ is the standard 
Bjorken variable for DIS off the proton and $\nu$ is the photon energy. 

A priori it is not clear that the nonlinear (\ref{eq:DSDT}) can be cast in the 
linear form (\ref{eq:DSDT2}). Furthermore, the expectation values of the
square of the dipole cross section in (\ref{eq:DSDT2}) tend to be
dominated by the contribution from large, non-perturbative, dipoles:
$r \sim r_f=1/m_f \sim 1$ fm for the excitation of $q\bar{q}$ states,
which is the Born term for $\beta \sim 1$, and  $\rho \sim R_c \sim 0.25$ fm
for the Born term of high-mass, i.e., small-$\beta$,
diffractive DIS: excitation of the $q\bar{q}g$ Fock states
(for the determination of the propagation radius of perturbative gluons, $R_c$,
from lattice QCD and elsewhere see \cite{MAGIO,Field}). 
Still, despite the manifestly nonperturbative Born term, the
resummation of double-leading-log approximation (DLLA) - 
strongly ordered energy and dipole size -
contributions from $q\bar{q}g_1...g_n$ excitation 
is possible and has been shown to correspond to the familiar DLLA evolution 
of the diffractive structure function \cite{NZ3P}. Starting from \cite{GNZ95DifDIS}, 
in the phenomenological studies of diffractive DIS it has become customary 
to apply the DGLAP evolution \cite{DGLAP} to the whole diffractive structure 
function (SF)
\bea
f^{D(4)}(t=0,x_{\Pom},\beta,Q^2)= x_{\Pom}F_2^{D(4)}(t=0,x_{\Pom},\beta,Q^2)
= {\frac{Q^2}{ 4\pi^2\alpha_{em}}}
(Q^{2}+M^{2})\left.{\frac{d\sigma^{D}} 
{ dt dM^{2}}}\right|_{t=0}
\label{eq:1.9}
\eea 
(being differential in $t$, the so-defined diffractive SF is dimensionfull, 
and $\sigma^{\Pom}(x_{\Pom},\beta,\br)$ has a dimension $[{\rm mb}]^2$,
but that
does not affect its evolution properties. The $t$-integrated diffractive SF
is dimensionless, but the modulation of the SF by the $\beta$-dependent 
diffraction slope \cite{NPZdifslope} can spoil the evolution properties, and
herebelow we focus on forward diffraction, $t=0$.) Although it has been 
argued to be plausible \cite{Collins}, and the DGLAP evolution analyses met certain 
phenomenological success (\cite{BartelsDifDIS,H1,ZEUS} and references therein,
for the review see \cite{Hebecker}), a direct demonstration of
such a  DGLAP evolution property of diffractive DIS is still missing. 

The principal problem with extension of the analysis \cite{NZ3P} to 
the contribution of the 
$(q\bar{q})(f\bar{f})$, $(q\bar{q})g_1...g_n(f\bar{f})$ Fock states is that 
the $f\bar{f}$ dipoles have a large non-perturbative size,
$R \sim r_f = 1/m_f \gg R_c$. The gross features of  
$\beta$-distribution in $\gamma^* p \to (f\bar{f})p'$ are well 
understood: in close analogy to the valence 
structure function of the proton, it is peaked at $\beta \sim 1/2$, 
so that only finite masses, $M^2\sim Q^2$, are excited
\cite{NZ92,GNZcharm}. However, it is not obvious that this
non-perturbative valence $\beta$-distribution defined by  
$\gamma^*p \to (f\bar{f})p'$ will
enter the evolution of the $q\bar{q}$ sea of the pomeron
in precisely the same way as the valence quark density 
enters the evolution of the sea of nucleons. Here we report a direct 
demonstration that such a pQCD
evolution holds at least to the DLLA accuracy. Furthermore, we show that
although the $(q\bar{q})(f\bar{f})$ contribution is of higher order
in the pQCD coupling $\alpha_S$, see Fig.~1,  it  
is enhanced by a potentially large numerical factor, 
$\propto \left[\sigma(r_f)/\sigma(R_c)\right]^2$, and numerically
it is comparable to
the leading order $q\bar{q}g$ contribution. We report also a derivation
of the inclusive spectrum of gluon jets from diffraction excitation 
of the $q\bar{q}g$ states of the photon, which clarifies the small-$\br^2$ 
scaling properties
of $\sigma^{\Pom}(x_{\Pom},\beta,\br)$.

The further presentation is organized as follows. We start with
the brief introduction into the color dipole description of
small-$\beta$ diffraction and demonstration of the representation
(\ref{eq:DSDT2}) for excitation of the $q\bar qg$ state. Then we 
show how the DLLA contribution from $(q\bar q)(f\bar f)$ excitation
to $f^{D(4)}(t=0,x_{\Pom},\beta,Q^2)$  can be cast in the form (\ref{eq:DSDT2}) with 
$\sigma^{\Pom}(x_{\Pom},\beta,\br)$ evaluated 
for scattering of the $q\bar{q}$ dipole
on the $f\bar{f}$ valence state of the pomeron. We present the 
DLLA evaluation of the small-$\beta$ diffractive SF, compare our 
results with experimental data \cite{H1,ZEUS}
and conclude with a brief summary.

In the color dipole QCD approach
to DIS \cite{NZ91,NZ3P,NZZ94}
the two principal quantities are the  dipole cross section,
$\sigma(x,\br)$, for interaction of
the $q\bar{q}$ dipole ${\bf r}$  with
the proton target 
and  the $q\bar{q}$ dipole size distribution 
in the projectile  photon
$e_q^2\vert\Psi_{\gamma^{*}}(Q^{2},z,\br)\vert^{2}$.
In terms of $\sigma(x,\br)$ the cross section of inclusive DIS has
the form of an expectation value
over the $q\bar{q}$ Fock state, 
$\sigma_{\gamma^*p}(x,Q^2)=\langle q\bar{q}|\sigma(x,\br)| q\bar{q}\rangle$, 
the  effect of higher order perturbative 
Fock states, $q\bar{q}g_1...g_n$,
can be reabsorbed into the leading $\log{(1/ x)}$  color dipole 
BFKL evolution of $\sigma(x,\br)$.  
The relationship between the dipole cross section and the unintegrated
gluon structure function
${{\cal F}(x,\kappa^2)}={\partial G(x,\kappa^2)/\partial\log{\kappa^2}},$ reads
\bea
\sigma(x,\br)={\frac{4\pi \alpha_S(r^2)}{N_c}}
\int{\frac{d^2\bkappa}{(\kappa^2+\mu_G^2)^2}}
[1-\exp(i\bkappa\br)]{\cal F}(x,\kappa^2) \approx {\frac{\pi^2\alpha_S(r^2)}
{ N_c}} r^2
G(x,{\frac{A}{ r^2}}),
\label{eq:2.3}
\eea
where $R_c=1/\mu_G$ is the Yukawa correlation radius for perturbative gluons
and in the DLLA for small dipoles  $A\simeq 10$.
Because of $\log Q^2$ scaling violations $G(x,{A/r^2})$  rises with 
the hard scale $A/\br^2$. To the lowest order in pQCD \cite{NZ91,NZsplit}
\beq
{\cal F}(x,\bkappa^2)= {\frac{C_F \alpha_S(\bkappa^2)}{\pi}} \cdot N_c V_N(\bkappa)\,,
\label{eq:2.4}
\eeq
where $N_c V_N(\bkappa)$ can be regarded as the number of
valence partons in the proton resolved by gluons at the scale $\bkappa^2$.
Here the vertex function $V_N(\bkappa)=1-G_2(\bkappa,-\bkappa)$ and the 
two-quark form-factor of the target nucleon, $G_2(\bkappa,-\bkappa)$,
vanishes for $\bkappa^2 \gsim R_N^{-2}$, where $R_N$ is the 
radius of the nucleon.

The $q\bar q g$ contribution to (\ref{eq:DSDT}) describes the first iteration
of 
the $\log{(1/\beta)}$ evolution of diffractive DIS and can be separated 
into the radiative correction to the small-mass $q\bar q$ excitation and 
the Born term of  the high-mass $q\bar q g$ excitation as follows.  
Let $\bf{r}$, ${\bfrho}$ and ${\bfrho}-{\bf r}$ be the $\bar{q}$-$q$, $g$-$q$ 
and $g$-$\bar{q}$ separations in the impact parameter (transverse size)
plane. The $q\bar{q}g$ 3-body interaction cross section equals \cite{NZ3P}
\beq
\sigma_{3}(x,\br,\brho)={\frac{C_A}{ 2C_F}}
[\sigma(x,\brho)+\sigma(x,\bfrho-{\bf r})] -
{\frac{1}{ N_c^2-1}}\sigma(x,\br)                         \, \, .
\label{eq:2.7}
\endeq
For soft perturbative gluons carrying a small fraction of photon's momentum,
$z_g\ll 1$, and  Yukawa infrared regularization 
the wave function of the 3-parton $q\bar{q}g$ state equals \cite{NZ3P,NZZ94}
\bea
&&|\Phi({\bf r},\bfrho,z,z_g)|^2=\nonumber\\
&&{\frac{e_q^2 C_F \alpha_S(r^2)}{\pi^2 z_g}}
|\Psi_{\gamma^*}(Q^2,z,\br)|^2
\left|\mu_{G}K_{1}(\mu_{G}\rho){\frac{\bfrho}{\rho}}
-\mu_{G}K_{1}(\mu_{G}|\bfrho-\bf r|)
{\frac{\bfrho-{\bf r}}{ |\bfrho-\bf r|}}\right|^{2}\nonumber\\
&&={ e_q^2 \over z_g}\cdot
|\Psi_{\gamma^*}(Q^2,z,\br)|^2 \cdot{\cal K}(\bfrho-{\bf r},\brho) \simeq 
 {e_q^2 C_F \alpha_S(r^2) \over \pi^2 z_g}
|\Psi_{\gamma^*}(Q^2,z,\br)|^2
{\frac{r^{2}}{\rho^{4}}}{ F}(\mu_{G}\rho)
\label{eq:2.8}
\eea
where ${\cal K}(\bfrho-{\bf r},\brho)$ is the kernel of the color dipole BFKL equation
\bea
{\partial \sigma(x,\br)\over \partial \log{1\over x}}=
{\cal K}\otimes \sigma(x,\br) &=&{2C_F \over C_A}\int d^2\brho\,\,
{\cal K}(\bfrho-{\bf r},\brho) [\sigma_{3}(x,\br,\brho)-\sigma(x,\br)]\nonumber\\ 
&=&
\int d^2\brho\,\,
{\cal K}(\bfrho-{\bf r},\brho) [\sigma(x,\brho)+\sigma(x,\brho-\br)-\sigma(x,\br)]
\label{eq:2.81}
\eea
and we also showed the DLLA approximation for $\br^2 \ll \brho^2$. The 
form factor of the infrared cutoff,
${ F}(z)$, satisfies ${ F}(0)=1$ and
${ F}(z) \propto\exp(-2z)$ at $z > 1$ \cite{NZ3P,GNZ95DifDIS}. 

Now notice, that in view of (\ref{eq:2.8}) the $q\bar{g}g$ 
contribution to(\ref{eq:DSDT}) can be rearranged as
\beq
{\cal K}\cdot (\sigma_3^2 -\sigma^2)= 2 \sigma {\cal K}\cdot (\sigma_3-\sigma)+
{\cal K}\cdot (\sigma_3-\sigma)^2\,. 
\label{eq:2.9}
\eeq
The first term in the rhs of (\ref{eq:2.9}) is the radiative correction 
to the small-mass $q\bar q$ excitation with the rapidity gap $x_{\Pom} \sim x$, 
i.e. the contribution from the 2-parton state to the total cross section
of diffraction (\ref{eq:DSDT}) must be 
calculated with the BFKL-evolved 
\beq
\sigma(x_{\Pom},\br)= \sigma(x_0,\br) + \int_{x_\Pom}^{x_{0}}{dx'\over x'} 
{\cal K}\otimes \sigma(x',\br)\,.
\label{eq:2.10}
\eeq 
The second term in (\ref{eq:2.9}) is the Born tern of 
the high-mass 3-parton, $q\bar{q}g$, 
excitation with the rapidity gap $x_{\Pom}=x_0$. In the high-mass regime
$dz_g/z_g = dM^2/(M^2+Q^2)$ and after undoing the $z_g$-integration the DLLA 
3-parton cross section takes the form 
\bea
&&(Q^{2}+M^{2})\left.{\frac{d\sigma_{q\bar{q}g}^{D}}{ dt dM^{2}}}\right|_{t=0}
= \nonumber\\
&&{\frac{1}{16\pi}}
\int dz d^{2}{\bf r}d^{2}{\bfrho}\,\,
\left\{z_{g}|\Phi({\bf r},{\bfrho},z,z_{g})|^{2}
\right\}_{z_{g}=0}
\cdot {\left[\sigma_{3}(x_{\Pom},\br,\brho)-
\sigma(x_{\Pom},\br)\right]^2 } \nonumber\\
\simeq &&{\frac{1}{16\pi}}\int dz \,d^{2}\br\,\,
|\Psi_{\gamma^{*}}(Q^{2},z,\br)|^{2}
{\frac{\pi^2 \alpha_{S}(r^2)}{ N_c}} r^{2} 
\cdot {C^2_A N_c\over C_F}
 \int_{r^2} d^2\brho
\left[{\frac{\sigma(x_{\Pom},\brho)}{\rho^{2}}}\right]^{2}
{ F}(\mu_G\rho) \,.
\label{eq:4.2}
\eea
It gives a flat small-$\beta$ behaviour of 
$f^{D(4)}(t=0,x_{\Pom},\beta,Q^2)$ with the strength 
controlled \cite{NZ3P,GNZcharm} by non-perturbative large $\rho$, 
cut off from above at $\rho \sim R_c$ by the nonperturbative 
form factor 
${ F}(\mu_{G}\rho)$. The small-$\rho$ integration can safely
be extended to $\rho=0$, so that (\ref{eq:4.2}) is of the 
desired color dipole form (\ref{eq:DSDT2}) and can be treated as
DIS off the sea generated by perturbative splitting of gluons from the 
valence $gg$ state of the pomeron. There is one caveat, though:
the gluon density in the pomeron defined by eq. ~(\ref{eq:4.2}),
\beq
G_{gg}^{\Pom}(x_{\Pom},\beta)={C^2_A N_c\over C_F}
\cdot \int d^2\brho
\left[{\frac{\sigma(x_{\Pom},\brho)}{ \rho^{2}}}\right]^{2}
{ F}(\mu_G\rho) \,,
\label{eq:4.21}
\eeq
is short of the collinear scaling violations present in (\ref{eq:2.3}).

The extension of the above analysis to the DLLA description of
diffractive excitation of the high-mass $(q\bar q)(f\bar f)$ Fock state of
the photon proceeds as follows.  As we shall see a posteriori,
the DLLA contribution comes from $\br^2 \ll \brho^2 \ll \bR^2$.
First we recall that the $q\bar{q}$ excitation is dominated by 
very asymmetric pairs, in which one of the final partons carries
a very small fraction of the photons
momentum, $z \sim m_q^2/Q^2 \ll 1$, so that in the impact parameter
space the fast parton with $\bar{z}=1-z$ 
flies along the photon's trajectory, while
the slower parton is at large transverse distance $r \sim 1/m_q$
from the parent photon \cite{NZ92}. Consequently, the fast parton
of the $f\bar{f}$ shall have the same impact parameter as the
gluon radiated by the parent $q\bar{q}$ dipole. In view of the
DLLA ordering, $\br^2 \ll \brho^2 \ll \bR^2$, the partons of the 
parent $q\bar{q}$ dipole and the fast parton of the radiative 
$f\bar{f}$ pair can be treated as the pointlike (anti)triplet color 
charge, and the $(q\bar q)(f\bar f)$ state interacts with the target
nucleon as the $f\bar f$ dipole with the dipole cross section
$\sigma(x_{\Pom},\bR)$. The distribution of $f\bar{f}$ color 
dipoles in the gluon of transverse momentum $\bkappa$ is 
identical to that in the photon subject to the substitutions
$N_c\alpha_{em} e_f^2\to T_F\alpha_S(\bkappa^2)$ and $Q^2 \to \bkappa^2$, 
so that the diffractive cross section of interest equals
\bea
(Q^{2}+M^{2})\left.{\frac{d\sigma^{D}_{(q\bar q)(f\bar f)}}{ dt dM^{2}}}\right|_{t=0}=
{1\over 16\pi}\int d^2\bkappa {\frac{dg_{q\bar{q}}(Q^2,\bkappa)}{ d^2\bkappa}}
\cdot {\frac{T_F\alpha_S(\bkappa^2)}{ N_c\alpha_{em} e_f^2}} 
\langle f\bar{f}|\sigma^2(x_{\Pom},\bR)|f\bar{f}\rangle\, ,
\label{eq:qqff}
\eea
where the flux of gluons in the parent $q\bar{q}$ state is given by the momentum-space version of
(\ref{eq:2.8}):
 \bea
{\frac{dg_{q\bar{q}}(Q^2,\bkappa)}{ d^2\bkappa}} =
 \int_0^1 dz_q
\int d^2{\bf r}\left|\Psi_{\gamma^*}(Q^2,z_q,{\bf r})\right|^2
 {\frac{2e_q^2C_F\alpha_S(r^2)}{ \pi^2}}\cdot 
{\frac{\left[1-\exp(i\bkappa{\bf r})\right]} 
{(\bkappa^2+\mu_G^2)^2}}\bkappa^2 \,.         
\label{eq:NG}
\eea
Finally, notice that 
\bea
{\frac{\bkappa^2}{ 4\pi^2 \alpha_{em}}}\cdot 
{\frac{1}{ 16\pi e_f^2}}\langle f\bar{f}|\sigma^2(x_{\Pom},\bR)|f\bar f\rangle &=&
{\frac{\bkappa^2}{ 4\pi^2 \alpha_{em}}}\cdot 
\left.{\frac{d\sigma(\gamma^*(\bkappa^2) \to f\bar{f})} { dt}}\right|_{t=0} \nonumber\\
= {\frac{1}{e_f^2}}
\int_{0}^{1} {\frac{d\beta} { \beta}} 
f^{D(4)}_{f\bar{f}}(t=0,x_{\Pom},\beta,\bkappa^2) &=& N_{f\bar{f}}^{\Pom}(x_{\Pom},\bkappa^2)
\label{eq:Nff}
\eea
where $N_{f\bar{f}}^{\Pom}(x_{\Pom},\bkappa^2)$ can be reinterpreted as a
number of charged valence partons, i.e., twice the number of $f\bar{f}$ dipoles,
in the pomeron. Upon the substitution of (\ref{eq:Nff}) and (\ref{eq:NG})
into (\ref{eq:qqff}) one readily recovers the dipole representation 
(\ref{eq:DSDT2}), in which $\sigma^{\Pom}(x_{\Pom},\beta,\br)$ is evaluated 
from equation (4) in which unintegrated gluon density (5) is substituted for
by the unintegrated gluon density evolved from the $f\bar{f}$ state of the
pomeron
\beq
{\cal F}_{f\bar{f}}^{\Pom}(\beta,\bkappa^2)= {\frac{C_F \alpha_S(\bkappa^2)}{ \pi}} 
N_{f\bar{f}}^{\Pom}(x_{\Pom},\bkappa^2)\, .
\label{eq:PomGlueff}
\eeq
Furthermore,  
$N_{f\bar{f}}^{\Pom}(x_{\Pom},\bkappa^2)$ vanishes at $\bkappa^2 =0$ and,
according to \cite{NZ91,NZ92}, 
flattens at $\bkappa^2 \gg m_f^2$ which, in comparison to (\ref{eq:2.4}) 
suggests the transverse size of the $f\bar{f}$ component of the pomeron
$r_f \sim 1/m_f$. One can come to the same conclusion from the point that
the dominant contribution to (\ref{eq:NG}) comes from $f\bar{f}$ dipoles
with $R \sim 1/m_f$.

The DLLA analysis of $q\bar{q}g_1...g_n$ excitation developed in \cite{NZ3P}
can readily be extended to the higher, $(q\bar{q})g_1..g_n(f\bar{f})$, states.
The crucial point is that to DLLA the $f\bar{f}$ dipole is the largest
one, so that the corresponding contribution to the diffractive cross
section is still given by equation (\ref{eq:qqff}) where the DLLA evolution
is reabsorbed into the flux of gluon $g_n$, which is the softest with respect
to the photon. Viewed from the pomeron side, that amounts to the DLLA
small-$\beta$ evolution of $\sigma^{\Pom}(x_{\Pom},\beta,\br)$ with the
boundary condition defined by gluon density (\ref{eq:PomGlueff}). As such, 
the emerging $\log^{n-1}{(1/\beta)} \cdot \log^n{(1/ \alpha_S(r^2))}$ 
structure of DLLA expansion in the energy and collinear logarithms 
for diffractive SF from $(q\bar{q})g_1..g_n(f\bar{f})$
excitation is identical to DLLA structure of the proton SF. As shown in
\cite{NZ3P},  DLLA expansion for diffractive SF from $(q\bar{q})g_1..g_n$ 
excitation is of a marginally different structure 
$\log^{n-1}{(1/ \beta)} \cdot \log^{n-1}{(1/\alpha_S(r^2))}$. Besides
that, the two components of the diffractive structure function have 
a manifestly different $x_{\Pom}$-dependence \cite{GNZ95DifDIS}: driven by 
$\sigma(x_{\Pom},r_f)$ in (\ref{eq:PomGlueff}) for the $q\bar{q},(q\bar{q})(f\bar{f}), 
(q\bar{q})g_1..g_n(f\bar{f})$ excitations and by $\sigma(x_{\Pom},R_c)$
for the $(q\bar{q})g_1..g_n$ excitations. This concludes the proof of 
the DLLA small-$\beta$ evolution at fixed $x_{\Pom}$ of such a
two-component diffractive 
structure function $f^{D(4)}(t=0,x_{\Pom},\beta,Q^2)$.

\begin{figure}[h]
\psfig{figure=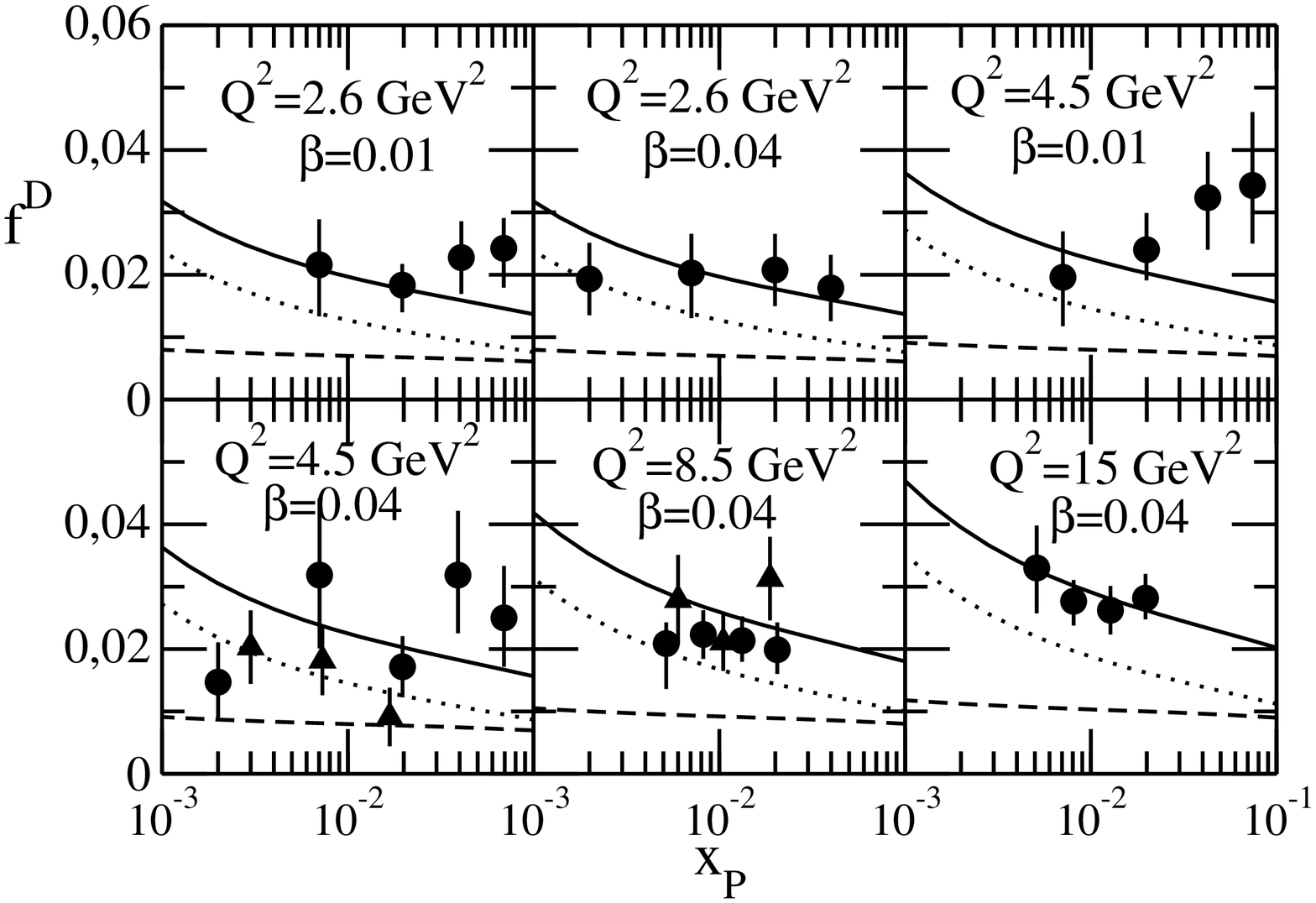,angle=-90,scale=0.6}
\vspace{-1.0cm}
\caption{The comparison with the experimental data 
on small-$\beta$, small-$x_{\Pom}$ diffractive structure function
(\protect\cite{H1}, full circles;
\protect\cite{ZEUS}, full triangle) 
of the theoretical evaluation of 
$f^{D(3)}=f_{q\bar qg}^{D(3)}+f_{(q\bar q)(f\bar f)}^{D(3)}$ shown by the solid line. 
The dotted line corresponds to $f^{D(3)}_{q\bar qg}$ and the dashed line
represents $f_{(q\bar q)(f\bar f)}^{D(3)}$.}
\label{fig:F3D}
\end{figure}  

The absence of a scaling violation in (\ref{eq:4.2})
and (\ref{eq:4.21}) implies that in contrast to (\ref{eq:2.4}) the 
corresponding unintegrated gluon density ${\cal F}_{gg}^D(\beta,\bkappa^2)$ 
vanishes for large $\bkappa^2$. A closely related observable is
the transverse momentum, $\bp$, spectrum of semihard diffractive gluons 
with $\bp^2 \ll \br^{-2} \ll Q^2 $ \cite{PomeronKt}. Since $\bp$ is a variable
conjugate to $\brho$, upon the relevant Fourier transforms
\bea
(Q^{2}+M^{2})\left.{\frac{d\sigma^{D}(\gamma^* \to g X)}{ dt dM^{2}d^2\bp} }\right|_{t=0}
=
{\frac {1} {16\pi(2\pi)^2}}\int dz d^{2}{\bf r}C_F|\Psi_{\gamma^*}(z,\br)|^2
{\frac{\alpha_S(r)}{\pi^2}}\nonumber\\
\times\left|\int d^{2}{\bfrho}\exp(i{\bf p}{\bfrho})
\left[{\frac{\bfrho}{\bfrho^2}}-{\frac {\bfrho-\bf r}{(\bfrho-\bf r)^2} }
\right]\left[\sigma_{3}(x_{\Pom},{r},\brho)-\sigma(x_{\Pom},\br)\right]\right|^2\nonumber\\
=\int dz d^{2}{\bf r}|\Psi_{\gamma^*}(z,\br)|^2
{\frac{\pi\alpha_S(r^2)}{ N_c}}r^2 \cdot {\frac{C_A^2}{4C_F N_c }}\cdot\left[ 
{\frac{
\alpha_S({\bf p}^2)G(x_{\Pom},{\bf p}^2)} { \bp^2}} \right]^2
\label{eq:3.5}
\eea
Within the reinterpretation of diffraction as DIS of pomerons, the $\bp$ 
has a meaning of intrinsic transverse momentum in the valence $gg$ state of
the pomeron. Indeed, the spectrum (\ref{eq:3.5}) falls steeper than the 
$1/ \bp^2$ spectrum of gluons from inclusive DIS off a nucleon.

The numerical results for high-mass, small-$\beta$,  
diffraction depend on the input dipole cross section $\sigma(x,\br)$.
Here we evaluate the lowest order $q\bar{q}g$ and $(q\bar{q})(f\bar{f})$
contributions to diffractive DIS in a specific  color 
dipole BFKL model \cite{NZZ94,CDBFKL} which gives a good description
of the proton SF data. The applicability domain of
the small-$\beta$, small-$x_{\Pom}$ formalism is $\beta,x_{\Pom} < x_0 \ll 1$,
the experience with inclusive DIS suggests $x_{0}\sim 0.03$, although
the theoretical curves in fig.~2 are stretched up to $x_{\Pom}=0.1$. 
This small-$\beta$, small-$x_{\Pom}$  
domain is almost at the boundary of the HERA experiments, the 
corresponding experimental data on the $t$-integrated diffractive structure function
$f^{D(3)}(x_{\Pom},\beta,Q^2)$ from H1 (\cite{H1}, circles) and ZEUS (\cite{ZEUS}, triangles)
are shown in fig.~2. We evaluate this structure function as
$f^{D(3)}(x_{\Pom},\beta,Q^2) = \int dt f^{D(4)}(t,x_{\Pom},\beta,Q^2) \approx
{\frac {1} {B_{3\Pom}}} f^{D(4)}(t=0,x_{\Pom},\beta,Q^2)$ with the central value of
the diffraction slope $B_{\Pom} =
B_D=7.2\pm 1.1^{+0.7}_{-0.9}$ GeV$^{-2}$ as reported by ZEUS 
\cite{BD}.  The apparent growth of the experimentally observed 
$f^{D(3)}(x_{\Pom},\beta,Q^2)$  towards large $x_{\Pom}\sim 0.1$ is usually
attributed to the non-vacuum admixture to the pomeron exchange.
Two features of the theoretical results for small-$\beta$ 
diffraction are noteworthy. First, the contributions from $q\bar{q}g$ and 
higher-order $(q\bar{q})(f\bar{f})$ states are of comparable magnitude
because $R_c \ll r_f$ and the latter is enhanced 
$\propto [\sigma(x_{\Pom},r_f)/\sigma(x_{\Pom},R_c)]^2$.
Second, because of the same inequality of the important dipole sizes,
$R_c \ll r_f$,
the $x_{\Pom}$-dependence of the $q\bar{q}g$ excitation is steeper 
than that of the  $(q\bar{q})(f\bar{f})$ excitation. This point has been
made already in \cite{GNZ95DifDIS}, the numerically significant contribution
from the $(q\bar{q})(f\bar{f})$ excitation makes the overall
$x_{\Pom}$-dependence of $f^{D(3)}(x_{\Pom},\beta,Q^2)$ weaker than evaluated
in \cite{GNZ95DifDIS} for the pure  $q\bar{q}g$ excitation. The solid curve
in fig.~2 is the combined contribution from the two mechanisms. It is in reasonably
good agreement with the HERA data.

To summarize, we reported the first explicit proof of the 
DLLA evolution property of the contribution to diffractive structure 
function from excitation of $(q\bar q)(f\bar f),(q\bar q)g_1...g_n(f\bar f)$
Fock states of the photon. We demonstrated that the corresponding 
diffractive SF can be cast in the color dipole representation. The
boundary condition for the DLLA small-$\beta$ evolution is provided
by the Born dipole cross section built perturbatively  
upon the valence $f\bar{f}$
state of the pomeron, as defined by the $\gamma^* p \to (f\bar{f}) p'$
excitation, in precisely the same manner as in inclusive DIS off the
nucleon starting with the valence quark distribution. Compared to the
$q\bar{q}g$ excitation, the $(q\bar q)(f\bar f)$ is of higher order
to pQCD. Still the numerical evaluations confirm the expectation that
the pQCD $\alpha_S$ suppression is compensated for by the larger dipoles in
the $(q\bar q)(f\bar f)$ state compared to the $q\bar{q}g$ state of
the photon. 

This work has been partly supported by the INTAS grant 00-00366 and
the DFG grant 436RUS17/72/03.

\end{document}